\documentclass{article}
\usepackage{amsmath}

\input{tcilatex}

\begin{document}

\title{Anisotropic evolution of a D-brane}
\author{Pawel Gusin \\
University of Silesia, Institute of Physics, ul. Uniwersytecka 4, \\
PL-40007 Katowice, Poland,\\
e-mail: pgusin@us.edu.pl }
\maketitle

\begin{abstract}
The evolution of a probe D-brane in the p-brane background has been
considered. The anisotropic evolution of the world-volume of the D-brane
with a given topology of a world-volume in a form of a direct product of a $n
$-dimensional flat space and ($3-n$)-dimensional sphere has been obtained.
In this case the anisotropy is described with the aid of two parameters
(Hubble parameters).\ The special case of this evolution, namely the
isotropic evolution corresponds to equality of these two parameters. In the
latter case the masses and charges of the background p-branes have been
derived.

\begin{description}
\item[PACS] : 11.25.-w ; 11.27.+d

\item[Keywords] : D-branes, p-branes, Hubble parameters
\end{description}
\end{abstract}

\section{Introduction}

The motion of the D-branes in the diversity backgrounds has been considered
in the variety of papers e.g. [1-5]. The applications D-branes to the
cosmology and gravity are also widely discussed e.g. [2-4]. In these
approaches a background is fixed by the solutions of the superstring theory
approximated by a ten dimensional supergravity. In type II superstring
theory the background fields in the bosonic sector are given by a ten
dimensional metric, a dilaton and an antisymmetric field $B$ which come from
the NS sector. The sector RR gives antisymmetric fields represented by
p-forms. In this approximation a Dp-brane is a p-dimensional submanifold
embedded in the background. Then the motion of the D-brane in the ten
dimensional spacetime is determined from the action which consists of
Dirac-Born-Infeld (DBI) term and WZW-like term. In the string theory the
D-brane is the place where open strings terminate. In the world-volume of
the D-brane the fields are realized by two mechanism. The first one is
related to the ends of the open strings and the second one is related to the
pull-backs of the background fields from NS and RR sectors to the
world-volume. The former realization gives the fields which in same cases
corresponds to the content the Standard Model (SM). Hence one can consider
the world-volume of the brane as a model of the universe. The second
mechanism is related to the realization gravity in the world-volume. The
graviton and other fields from NS and RR sectors are not confined to the
world-volume (as the fields of SM) but propagate in the extra dimensions
perpendicular to the brane. These extra dimensions can have big sizes e.
g.[6-9] or their sizes are compact and small. In the last case one has to
compactificate six spatial dimmensions. The extra dimensions warp on cycles
of a\emph{\ }compact internal manifold. The topology and the geometry of
this internal manifold is constrained by the consistency conditions such as
a supersymmetry. In the case when fluxes induced by the fields from NS and
RR sectors vanish and supersymmetry is $N=2$ in four dimensional Minkowski
spacetime, then the internal manifold is Calabi-Yau. Non-vanishing fluxes
break supersymmetry and the internal manifold is deformed so the Calabi-Yau
manifold is replaced by the generalized complex manifold [10-14].

The D-brane is considered as an embedded submanifold in the ten-dimensional
spacetime with nontrivial backgrounds fields. The effective theory is
described by the Dirac-Born-Infeld (DBI) action, which determine the motion
in the spacetime. From the D-brane point of view this motion is interpreted
as the evolution of the world-volume of the brane. In the D-brane models of
the universe all fields and particles of the Standard Model are fixed to the
world-volume. Thus evolution of D-brane corresponds to cosmological
evolution for the observer fixed to the world-volume.

The aim of this paper is determination of the backgrounds and their
parameters which are compatible with the observed isotropy of expansion of
the universe. This condition is expressed by the Hubble parameters. We
consider the cosmological evolution of a probe Dk-brane with the DBI action
in the backgrounds of p-branes. The probe Dk-brane means that the
backreactions are neglected. We assume that the Dk-brane has the topology of
the direct product of two spaces: one is flat and the second has a constant
curvature. In this case we will obtain the Hubble parameters for these
spaces. These parameters are depend on the tangent and the normal directions
to Dk-brane. The condition for the equality of these parameters realize
isotropy of the expansion.

In the section 2 we recall the DBI action for a Dk-brane in the backgrounds
produced by p-branes and derive it equation of motion in given embedding. In
the section 3 we derive a ratio of the Hubble parameters for k=3. These
parameters are obtained from the metric on the world-volume of D3-brane.
This case (k=3) can be considered as a toy cosmological model corresponding
to 3+1 spacetime. The cosmological models derived from M-theory admit warped
factors which depend on time [15]. On the type IIB string theory side these
factors correspond to different rates expansion of the tangent directions to
a D3-brane. The section 4 is devoted to conclusions.

\section{D-brane evolution}

In this section we consider the motion of a Dk-brane. The Dk-brane action is
described by the DBI action:%
\begin{equation}
S=-T_{k}\int d^{k+1}\xi e^{-\phi }\sqrt{-\det (\gamma _{\mu \nu }+2\pi
\alpha ^{\prime }F_{\mu \nu }+B_{\mu \nu })}+T_{k}\int \sum_{i}\widetilde{A}%
_{\left( i\right) }\wedge e^{2\pi \alpha ^{\prime }F+B},  \tag{2.1}
\end{equation}%
where $\gamma _{\mu \nu }$ is the pull back of the background metric, $%
B_{\mu \nu }$ is the pull back of the background the NS 2-form, $F_{\mu \nu
} $ is the strength of the abelian gauge field on the worldvolume and $%
\widetilde{A}_{\left( i\right) }$ are pull-back of the background $i$-forms $%
A_{\left( i\right) }$ with odd (even) degrees: $i=1,3,5,7$ ($i=0,2,4,6,8$)
in the Type IIA (IIB) theory. We consider the background solutions with the
symmetry group $\mathbf{R}^{1}\times E_{\left( 6-p\right) }\times SO\left(
p+3\right) $, where $E_{\left( 6-p\right) }$ is the Euclidean group. They
are given by [16, 17, 18, 19]:

\begin{itemize}
\item the metric:%
\begin{equation}
ds^{2}=-\Delta _{+}\Delta _{-}^{-\frac{7-p}{8}}dt^{2}+\Delta _{+}^{-1}\Delta
_{-}^{\frac{\left( 3-p\right) ^{2}}{2\left( 1+p\right) }-1}dr^{2}+r^{2}%
\Delta _{-}^{\frac{\left( 3-p\right) ^{2}}{2\left( 1+p\right) }}d\Omega
_{p+2}^{2}+\Delta _{-}^{\frac{1+p}{8}}dX_{i}dX^{i},  \tag{2.2}
\end{equation}%
where:%
\begin{equation*}
\Delta _{\pm }\left( r\right) =1-\left( \frac{r_{\pm }}{r}\right) ^{p+1},
\end{equation*}%
$d\Omega _{p+2}^{2}$ is the metric on a (p+2)-dimensional sphere $S^{p+2}$ :%
\begin{equation*}
d\Omega _{p+2}^{2}=h_{rs}d\theta ^{r}d\theta ^{s},
\end{equation*}%
$r,s=1,...,p+2$ and $i=1,...6-p$,

\item the gauge strength $F=dA_{\left( p+1\right) }$ :%
\begin{equation}
F=\left( p+1\right) \left( r_{+}r_{-}\right) ^{\left( p+1\right)
/2}\varepsilon _{p+2},  \tag{2.3 }
\end{equation}%
$\varepsilon _{p+2}$ is the volume form on $S^{p+2}$,

\item the dilaton field:%
\begin{equation}
e^{-2\phi }=\Delta _{-}^{a},  \tag{2.4 }
\end{equation}%
where $a^{2}=\left( 3-p\right) ^{2}/4$.
\end{itemize}

The topological charge $g_{6-p}$ and the mass $m_{6-p}$ of the background
are expressed by $r_{+}$, $r_{-}$ :%
\begin{equation}
g_{6-p}=\frac{vol\left( S^{p+2}\right) }{\sqrt{2}\kappa }\left( p+1\right)
\left( r_{+}r_{-}\right) ^{\left( p+1\right) /2},  \tag{2.5}
\end{equation}%
\begin{equation}
m_{6-p}=\frac{vol\left( S^{p+2}\right) }{2\kappa ^{2}}\left[ \left(
p+2\right) r_{+}^{p+1}-r_{-}^{p+1}\right] .  \tag{2.6}
\end{equation}%
The above solution becomes the BPS state for $r_{+}=r_{-}=R$ with the metric:%
\begin{equation}
ds^{2}=\Delta ^{\frac{p+1}{8}}\left( -dt^{2}+dX_{i}dX^{i}\right) +\Delta ^{%
\frac{p-7}{8}}\left( d\rho ^{2}+\rho ^{2}d\Omega _{p+2}^{2}\right) , 
\tag{2.7}
\end{equation}%
where $\rho $ is related to $r$ as follows: $\rho ^{p+1}=r^{p+1}-R^{p+1}$
and $\Delta =1+\left( R/\rho \right) ^{p+1}$.

We have considered this background solution since they are general and for $%
p=3$ the last metric has the form used in the warp compactification.

In the general case the Dk-brane and D(6-p)-brane do not intersect if their
spatial dimensions obey the relation:%
\begin{equation*}
6\geq k+6-p.
\end{equation*}%
We denote the background coordinates as follows:%
\begin{equation*}
X^{M}=(t,X^{1},...,X^{6-p},r,\varphi ^{1},...,\varphi ^{p+2}),
\end{equation*}%
where $\varphi ^{1},...,\varphi ^{p+2}$ are coordinates on the sphere $%
S^{p+2}$, so $r$ and $\varphi ^{1},...,\varphi ^{p+2}$ span the transverse
directions to the (6-p)-brane. The Dk-brane propagating in this background
has $n$-directions parallel to (6-p)-brane and $k-n$ directions
perpendicular to (6-p)-brane where the number $n$ is given by [1]:%
\begin{equation}
n\leq n_{0}=\min (k,6-p).  \tag{2.8}
\end{equation}%
We will consider free falling Dk-brane in its rest frame with the proper
time $\tau $. We assume that $r$ is always transverse to Dk-brane and
Dk-brane has the topology of the direct product:%
\begin{equation*}
V_{n}\times S^{k-n},
\end{equation*}%
where $V_{n}$ is some n-dimensional flat space and $S^{k-n}$ is
(k-n)-dimensional sphere. Thus the embedding field has the form:%
\begin{equation}
X^{M}\left( \tau \right) =\left( t\left( \tau \right) ,\xi ^{1},...,\xi
^{n},X^{n+1},...,X^{6-p},r\left( \tau \right) ,\theta ^{1},...,\theta
^{k-n},\varphi ^{k-n+1}\left( \tau \right) ,...,\varphi ^{p+2}\left( \tau
\right) \right) ,  \tag{2.9}
\end{equation}%
where $\xi ^{1},...,\xi ^{n}$ are coordinates on $V_{n}$ and $\theta
^{1},...,\theta ^{k-n}$ are coordinates on $S^{k-n}$. The induced metric $%
\gamma _{\mu \nu }$ on the world-volume by the embedding (2.9) has the form:%
\begin{equation}
\gamma _{00}=-\Delta _{+}\Delta _{-}^{-\frac{7-p}{8}}\overset{\cdot }{t}%
^{2}+\Delta _{+}^{-1}\Delta _{-}^{\frac{\left( 3-p\right) ^{2}}{2\left(
1+p\right) }-1}\overset{\cdot }{r}^{2}+r^{2}h_{\widehat{r}\widehat{s}}%
\overset{\cdot }{\varphi }^{\widehat{r}}\overset{\cdot }{\varphi }^{\widehat{%
s}},  \tag{2.10}
\end{equation}%
\begin{equation}
\gamma _{ab}=\Delta _{-}^{\frac{1+p}{8}}\delta _{ab},  \tag{2.11}
\end{equation}%
\begin{equation}
\gamma _{\widehat{a}\widehat{b}}=r^{2}h_{\widehat{a}\widehat{b}},  \tag{2.12}
\end{equation}%
where $\widehat{r}$, $\widehat{s}=k-n+1,...,p+2$ , $a,b=1,...n$, and $%
\widehat{a},\widehat{b}=1,...,k-n$. The metrics $h_{\widehat{a}\widehat{b}}$%
\ and $h_{\widehat{r}\widehat{s}}$ are expressed by the metric $h_{rs}$ on
the sphere $S^{p+2}$:%
\begin{equation*}
\left( h_{rs}\right) =\left( 
\begin{array}{cc}
\left( h_{\widehat{a}\widehat{b}}\right) &  \\ 
& \left( h_{\widehat{r}\widehat{s}}\right)%
\end{array}%
\right) .
\end{equation*}%
The dot over coordinates means the derivative with the respect to the proper
time $\tau $. In the case when the background NS form $B$ is zero and the
abelian gauge field on the worldvolume vanishes the WZ term in (2.1) takes
the form:%
\begin{equation*}
\int \widetilde{A}_{\left( k+1\right) },
\end{equation*}%
where the form $\widetilde{A}_{\left( k+1\right) }$ is given by the pull
back of background form $A_{\left( k+1\right) }$. In the considered
background the only non-vanishing form is $A_{\left( p+1\right)
}=A_{M_{0}...M_{p}}dX^{M_{0}}\wedge ...dX^{M_{p}}$ such that $dA_{\left(
p+1\right) }$ is given by (2.3). Thus WZ term does not vanish if k=p and the
DBI action takes the form:

\begin{equation}
S=-T_{k}\int d\tau d^{n}\xi d^{k-n}\theta \left( e^{-\phi }\sqrt{-\det
\left( \gamma _{\mu \nu }\right) }-\delta _{k,p}A\overset{\cdot }{t}\right) 
\tag{2.13}
\end{equation}%
and $A=A_{0...p}$. Since the terms in (2.13) do not depend on coordinates $%
\xi $ we get:%
\begin{equation*}
S=-T_{k}vol\left( V_{n}\right) \int d\tau d^{k-n}\theta \left( e^{-\phi }%
\sqrt{-\det \left( \gamma _{\mu \nu }\right) }-\delta _{k,p}A\overset{\cdot }%
{t}\right) .
\end{equation*}%
\ In the considered background we obtain:%
\begin{gather}
e^{-\phi }\sqrt{-\det \left( \gamma _{\mu \nu }\right) }=\left( \overset{%
\cdot }{t}^{2}-\Delta _{+}^{-2}\Delta _{-}^{\frac{1-p}{1+p}}\overset{\cdot }{%
r}^{2}-\Delta _{+}^{-1}\Delta _{-}^{\frac{7-p}{8}}r^{2}h_{\widehat{r}%
\widehat{s}}\overset{\cdot }{\varphi }^{\widehat{r}}\overset{\cdot }{\varphi 
}^{\widehat{s}}\right) ^{1/2}\times  \notag \\
r^{k-n}\Delta _{+}^{1/2}\Delta _{-}^{[5\left( 1-p\right) +n\left( 1+p\right)
]/16}\sqrt{\det \left( h_{\widehat{a}\widehat{b}}\right) }.  \tag{2.14}
\end{gather}%
In non-rotating case $\overset{\cdot }{\varphi }^{\widehat{r}}=0$ the action
simplifies and takes the form:%
\begin{equation}
S=-T_{k}vol\left( V_{n}\right) \int d\tau L,  \tag{2.15}
\end{equation}%
where the Lagrangian $L$ has the form:%
\begin{equation}
L=vol\left( S^{k-n}\right) \left( \overset{\cdot }{t}^{2}-\Delta
_{+}^{-2}\Delta _{-}^{\frac{1-p}{1+p}}\overset{\cdot }{r}^{2}\right)
^{1/2}r^{k-n}\Delta _{+}^{1/2}\Delta _{-}^{[5\left( 1-p\right) +n\left(
1+p\right) ]/16}-\delta _{k,p}Aw\overset{\cdot }{t}  \tag{2.16}
\end{equation}%
and $w=\int d^{k-n}\theta $ . Variation $L$ with respect to $t$ gives:%
\begin{equation}
vol\left( S^{k-n}\right) \frac{\overset{\cdot }{t}r^{k-n}\Delta
_{+}^{1/2}\Delta _{-}^{[5\left( 1-p\right) +n\left( 1+p\right) ]/16}}{\sqrt{%
\overset{\cdot }{t}^{2}-\Delta _{+}^{-2}\Delta _{-}^{\frac{1-p}{1+p}}\overset%
{\cdot }{r}^{2}}}-\delta _{k,p}Aw=E,  \tag{2.17}
\end{equation}%
where $E$ is a constant of motion.\ Thus:%
\begin{equation}
\left( \frac{dr}{dt}\right) ^{2}=\left[ 1-\frac{r^{2\left( k-n\right)
}\Delta _{+}^{1/2}\Delta _{-}^{[5\left( 1-p\right) +n\left( 1+p\right) ]/16}%
}{\left( E+\delta _{k,p}Aw\right) ^{2}}vol^{2}\left( S^{k-n}\right) \right]
\Delta _{+}^{2}\Delta _{-}^{-\frac{1-p}{1+p}}.  \tag{2.18}
\end{equation}%
The proper time $\tau $\ of the Dk-brane\ is expressed by:%
\begin{equation*}
d\tau ^{2}=\gamma _{\mu \nu }d\xi ^{\mu }d\xi ^{\nu }=g_{MN}\partial _{\mu
}X^{M}\partial _{\nu }X^{N}d\xi ^{\mu }d\xi ^{\nu }.
\end{equation*}%
In the rest frame of the Dk-brane and for the considering embedding this
proper time has the form:%
\begin{equation}
d\tau ^{2}=-\left( g_{00}+g_{rr}\overset{\cdot }{r}^{2}+r^{2}h_{\widehat{r}%
\widehat{s}}\overset{\cdot }{\varphi }^{\widehat{r}}\overset{\cdot }{\varphi 
}^{\widehat{s}}\right) dt^{2},  \tag{2.19}
\end{equation}%
where:%
\begin{eqnarray*}
g_{00} &=&-\Delta _{+}\Delta _{-}^{-\frac{7-p}{8}}, \\
g_{rr} &=&\Delta _{+}^{-1}\Delta _{-}^{\frac{\left( 3-p\right) ^{2}}{2\left(
1+p\right) }-1}.
\end{eqnarray*}%
In the non-rotating case $\overset{\cdot }{\varphi }^{\widehat{r}}=0$ the
derivatives with the respect to the proper time $\tau $ and coordinate time $%
t$ are related:%
\begin{equation*}
\left( \frac{dr}{dt}\right) ^{2}=\left( \frac{dr}{d\tau }\right) ^{2}\left( 
\frac{d\tau }{dt}\right) ^{2},
\end{equation*}%
so:%
\begin{equation*}
\left( \frac{dr}{dt}\right) ^{2}=-\frac{g_{00}}{1+g_{rr}\left( \frac{dr}{%
d\tau }\right) ^{2}}\left( \frac{dr}{d\tau }\right) ^{2}.
\end{equation*}%
From (2.18) and (2.19) we obtain relation between the radial position and
the proper time:%
\begin{equation}
\left( \frac{dr}{d\tau }\right) ^{2}=\frac{\left( E-\delta _{k,p}Aw\right)
^{2}-r^{k-n}\Delta _{+}^{1/2}\Delta _{-}^{\beta }vol^{2}\left(
S^{k-n}\right) }{\left( E-\delta _{k,p}Aw\right) ^{2}-\Delta _{-}^{\gamma
}+r^{k-n}\Delta _{+}^{3/2}\Delta _{-}^{\delta }vol^{2}\left( S^{k-n}\right) }%
\Delta _{+}\Delta _{-}^{\alpha },  \tag{2.20}
\end{equation}%
where the exponents are:%
\begin{equation*}
\alpha =\frac{-1+14p-p^{2}}{8\left( 1+p\right) },\text{ }\beta =\frac{%
5\left( 1-p\right) +n\left( 1+p\right) }{16},
\end{equation*}%
\begin{equation*}
\gamma =\frac{3\left( p^{2}-10p+9\right) }{8\left( 1+p\right) },\text{ }%
\delta =\frac{p^{2}-60p+61+n\left( 1+p\right) ^{2}}{16\left( 1+p\right) }.
\end{equation*}%
The equation (2.20) has the sense if the right side is greater or equal to
zero.

In the coordinate time $t$ the induced metric on the Dk-brane by the
embedding (2.9) has the form:%
\begin{equation}
ds^{2}=-\left( \Delta _{+}\Delta _{-}^{-\frac{7-p}{8}}-\Delta
_{+}^{-1}\Delta _{-}^{\frac{\left( 3-p\right) ^{2}}{2\left( 1+p\right) }-1}%
\overset{\cdot }{r}^{2}-r^{2}h_{\widehat{r}\widehat{s}}\overset{\cdot }{%
\varphi }^{\widehat{r}}\overset{\cdot }{\varphi }^{\widehat{s}}\right)
dt^{2}+\Delta _{-}^{\frac{1+p}{8}}d\xi _{a}d\xi ^{a}+r^{2}h_{\widehat{a}%
\widehat{b}}d\theta ^{\widehat{a}}d\theta ^{\widehat{b}}.  \tag{2.21}
\end{equation}%
Using (2.19) we get:%
\begin{equation}
ds^{2}=-d\tau ^{2}+e^{2\lambda }d\xi _{a}d\xi ^{a}+e^{2\beta }h_{\widehat{a}%
\widehat{b}}d\theta ^{\widehat{a}}d\theta ^{\widehat{b}},  \tag{2.22}
\end{equation}%
where $r\left( \tau \right) $ is the solution of the (2.20) and $\exp
2\lambda =\Delta _{-}^{\frac{1+p}{8}}$, $\exp 2\beta =r^{2}$. This metric
has the form of the Bianchi type for the homogenous space with two scale
factors namely $\Delta _{-}^{\frac{1+p}{8}}$ and $r^{2}$.

In the conformal time $x^{0}$ (which is related to $\tau $ as follows $%
dx^{0}=\exp \left( -\lambda \right) d\tau $) the metric (2.22) reads:%
\begin{equation}
ds^{2}=e^{2\lambda }\left( -\left( dx^{0}\right) ^{2}+d\xi _{a}d\xi
^{a}\right) +e^{2\beta }h_{\widehat{a}\widehat{b}}d\theta ^{\widehat{a}%
}d\theta ^{\widehat{b}}.  \tag{2.22a}
\end{equation}%
Then assuming that the evolution of the world-volume seeing by the observer
fixed to the brane, is governed by the Einstein gravity, the scale factor $%
\lambda $ and $\beta $ evolve as follow:%
\begin{equation}
n\left( n-1\right) \left( \lambda ^{\prime }\right) ^{2}+2mn\lambda ^{\prime
}\beta ^{\prime }+m\left( m-1\right) \left( \beta ^{\prime }\right)
^{2}+e^{2\lambda -2\beta }\widetilde{R}=16\pi G\widetilde{T}_{00}, 
\tag{2.23}
\end{equation}%
\begin{gather}
\left[ 2\left( 1-m\right) \lambda ^{\prime \prime }-2n\beta ^{\prime \prime
}+2n\left( 2-m\right) \lambda ^{\prime }\beta ^{\prime }+\left(
2n+m-m^{2}-2\right) \left( \lambda ^{\prime }\right) ^{2}\right.  \notag \\
\left. -n\left( n+1\right) \left( \beta ^{\prime }\right) ^{2}-e^{2\lambda
-2\beta }\widetilde{R}\right] \delta _{ab}=16\pi G\widetilde{T}_{ab}, 
\tag{2.24}
\end{gather}%
\begin{gather}
e^{2\beta -2\lambda }\left[ 2\left( 1-n\right) \beta ^{\prime \prime
}-2m\lambda ^{\prime \prime }+\left( 2m-mn-n\right) \lambda ^{\prime }\beta
^{\prime }-m\left( m-1\right) \left( \lambda ^{\prime }\right) ^{2}-n\left(
n-1\right) \left( \beta ^{\prime }\right) ^{2}\right] h_{\widehat{a}\widehat{%
b}}+  \notag \\
2\left( \widetilde{R}_{\widehat{a}\widehat{b}}-\frac{1}{2}\widetilde{R}h_{%
\widehat{a}\widehat{b}}\right) =16\pi G\widetilde{T}_{\widehat{a}\widehat{b}%
},  \tag{2.25}
\end{gather}%
where $m=k-n$, the Ricci tensor $\widetilde{R}_{\widehat{a}\widehat{b}}$ and
the scalar curvature $\widetilde{R}$ are obtained from the metric $h_{%
\widehat{a}\widehat{b}}$. The prime means derivative with respect to $x^{0}$
(e.g. : $\lambda ^{\prime }=d\lambda /dx^{0}$). The energy-momentum tensor $%
\widetilde{T}_{\mu \nu }=\left( \widetilde{T}_{00},\widetilde{T}_{ab},%
\widetilde{T}_{\widehat{a}\widehat{b}}\right) $ is given by the matter and
the fields in the world-volume of the Dk-brane and is defined with respect
to the metric (2.22a):%
\begin{equation}
\widetilde{T}_{\mu \nu }=\frac{2}{\sqrt{-g}}\frac{\partial \mathcal{L}_{m}}{%
\partial g^{\mu \nu }},  \tag{2.26}
\end{equation}%
where $\mathcal{L}_{m}$ is a Lagrangian density for the matter and the
fields in the world-volume. Using relations $\lambda ^{\prime }=\exp \left(
\lambda \right) d\lambda /d\tau $ and $\beta ^{\prime }=\exp \left( \lambda
\right) d\beta /d\tau $ one can rewrite eq. (2.23) as follows:%
\begin{equation}
n\left( n-1\right) H_{1}^{2}+2mnH_{1}H_{2}+m\left( m-1\right)
H_{2}^{2}+e^{-2\beta }\widetilde{R}=16\pi GT_{00},  \tag{2.27}
\end{equation}%
where $T_{00}$ is the energy density in the metric (2.22) related to $%
\widetilde{T}_{00}$ as follows: $\widetilde{T}_{00}e^{-2\lambda }=T_{00}$.
The quantities $H_{1}$ and $H_{2}$ are Hubble parameters given by: $%
H_{1}=d\lambda /d\tau $ and $H_{2}=d\beta /d\tau $. Hence from the eq.
(2.22) follows that the factor $\exp \left( 2\lambda \right) =\Delta
_{-}^{\left( 1+p\right) /8}$ correspods to the evolution in the flat
directions $\xi $ while the second factor $h_{\widehat{a}\widehat{b}}\exp
2\beta =r^{2}h_{\widehat{a}\widehat{b}}$ concerns the evolution in the
curved directions (corresponding to the sphere).

\section{Hubble parameters}

We have related the metric (2.22) to the two Hubble parameters:%
\begin{equation}
H_{1}=\frac{1}{\Delta _{-}^{\frac{1+p}{16}}}\frac{d}{d\tau }\left( \Delta
_{-}^{\frac{1+p}{16}}\right) ,  \tag{3.1}
\end{equation}%
\begin{equation}
H_{2}=\frac{1}{r}\frac{dr}{d\tau },  \tag{3.2}
\end{equation}%
where in $H_{2}$ is assumed isotropic evolution which means that $d\left( h_{%
\widehat{a}\widehat{b}}\right) /d\tau =0$. Then the eq. (3.1) reads:%
\begin{equation}
H_{1}=\frac{\left( p+1\right) ^{2}}{16}\cdot \frac{rr_{-}^{p+1}}{%
r^{p+1}-r_{-}^{p+1}}\frac{dr}{d\tau },  \tag{3.3}
\end{equation}%
where $dr/d\tau $ is given by (2.20). Thus the ratio of these Hubble
parameters is given by the relation:%
\begin{equation}
\frac{H_{1}}{H_{2}}=\frac{\left( p+1\right) ^{2}}{16}\cdot \frac{%
r^{2}r_{-}^{p+1}}{r^{p+1}-r_{-}^{p+1}}\equiv \eta \left( r\right) . 
\tag{3.4}
\end{equation}%
It depends on the position $r$ of the Dk-brane and $r$ is given by the
solution of the equation (2.20). In the case when $r_{+}=r_{-}=R$ and\ the
metric on the world-volume is given by (2.22) the eq. (2.18) takes the form:%
\begin{equation}
\left( \frac{dr}{dt}\right) ^{2}=\left[ 1-\frac{r^{2\left( k-n\right)
}\Delta ^{1/2+[5\left( 1-p\right) +n\left( 1+p\right) ]/16}}{\left( E+\delta
_{k,p}Aw\right) ^{2}}vol^{2}\left( S^{k-n}\right) \right] \Delta ^{\frac{1+3p%
}{1+p}}.  \tag{3.5}
\end{equation}%
The equation (3.5) can be interpreted as a one dimensional particle motion
with the zero energy in the potential $-V\left( r\right) $:%
\begin{equation*}
V\left( r\right) =\left[ 1-\frac{r^{2\left( k-n\right) }\Delta
^{1/2+[5\left( 1-p\right) +n\left( 1+p\right) ]/16}}{\left( E+\delta
_{k,p}Aw\right) ^{2}}vol^{2}\left( S^{k-n}\right) \right] \Delta ^{\frac{1+3p%
}{1+p}}.
\end{equation*}%
Hence $V$\ has to be positive and zeros of $V$\ gives the turning points.
The ratio of the Hubble parameters in this case is:%
\begin{equation}
\frac{H_{1}}{H_{2}}=\frac{\left( p+1\right) ^{2}}{16}\cdot \frac{r^{2}R^{p+1}%
}{r^{p+1}-R^{p+1}}.  \tag{3.6}
\end{equation}%
We restrict ourselves to the case when $k=3\ $which corresponds to the
D3-brane. In this case $a,b=1,...n$, and $\widehat{a},\widehat{b}=1,...,3-n$%
. From the other side the relation between the parameters $H_{1}$ and $H_{2}$
is given by the eq. (2.27). Thus in the conisdered case we obtain the
following equations:

for $n=0$:%
\begin{equation}
H_{2}^{2}+\frac{1}{6}e^{-2\beta }\widetilde{R}=\frac{8\pi G}{3}T_{00}, 
\tag{3.7}
\end{equation}

for $n=1$%
\begin{equation}
H_{2}^{2}+2H_{1}H_{2}+\frac{1}{2}e^{-2\beta }\widetilde{R}=8\pi GT_{00}, 
\tag{3.8}
\end{equation}

for $n=2$ 
\begin{equation}
H_{1}^{2}+2H_{1}H_{2}+\frac{1}{2}e^{-2\beta }\widetilde{R}=8\pi GT_{00}, 
\tag{3.9}
\end{equation}

for $n=3$%
\begin{equation}
H_{1}^{2}=\frac{8\pi G}{3}T_{00}.  \tag{3.10}
\end{equation}%
The eqs. (3.7) and (3.10) are equations only for $\beta $ and $\lambda $,
respectively. These equations govern the isotropic evolution of the
world-volume of the D3-brane. While the eqs. (3.8) and (3.9) give relation
between $H_{1}$ and $H_{2}$. One can see that the cases $n=1$ and $n=2$ are
symmetric, so we focus on the case $n=1$. Using the equation (3.4) the
equation (3.8) takes the form:%
\begin{equation}
H_{2}^{2}\left( 1+2\eta \right) +\frac{1}{2}e^{-2\beta }\widetilde{R}=8\pi
GT_{00}.  \tag{3.11}
\end{equation}
The all ingredients in the above equation are the function of $r$ which
evolve with respect to the coordinate time $t$ according to the eq.(3.5)
which, in the considered case ($k=3$), takes the form:%
\begin{equation}
\left( \frac{dr}{dt}\right) ^{2}=\left[ 1-\frac{r^{2\left( 3-n\right)
}\Delta ^{1/2+[5\left( 1-p\right) +n\left( 1+p\right) ]/16}}{\left( E+\delta
_{3,p}Aw\right) ^{2}}vol^{2}\left( S^{3-n}\right) \right] \Delta ^{\frac{1+3p%
}{1+p}}.  \tag{3.12}
\end{equation}%
In order to get how change (3.6) in the time $t$ we would like to solve the
eq. (3.12). These solutions among other depends on the dimension $p$ of the
background branes. Thus we have to consider each dimension $p$ separately.
Although in the type IIB the dimensions $p$ are odd we also consider the
background produced by D-particle, thus $p=0,1,3$ and $5$. The solutions of
(3.12) for differend $p$ are given below where the number of the non-compact
dimensions $n$ is given by the condition (2.8).

For $p=0$ (D-particle) the eq. (3.12) gives following result:%
\begin{equation}
\int \frac{\sqrt{r}dr}{\sqrt{r-R}\sqrt{1-r^{2\left( \alpha -\beta
_{0}\right) }\left( r-R\right) ^{2\beta _{0}}\sigma _{n}^{2}}}=t+t_{0}, 
\tag{3.13}
\end{equation}%
where%
\begin{equation*}
\sigma _{n}^{2}=\left( \frac{vol\left( S^{3-n}\right) }{E}\right) ^{2}
\end{equation*}%
and $\alpha =3-n$ , $2\beta _{0}=\left( 13+n\right) /16$. The number $n$ of
the flat dimensions is:%
\begin{equation*}
n=0,1,2,3.
\end{equation*}%
The cases $n=0$ and $n=3$ correspond to the only one Hubble parameter.

For $p=1$ (D-string) we get:%
\begin{equation}
\int \frac{r^{2}dr}{\left( r^{2}-R^{2}\right) \sqrt{1-r^{2\left( \alpha
-\beta _{1}\right) }\left( r^{2}-R^{2}\right) ^{2\beta _{1}}\sigma _{n}^{2}}}%
=t+t_{0},  \tag{3.14}
\end{equation}%
where $2\beta _{1}=\left( 9+2n\right) /16$.

For $p=3$:%
\begin{equation}
\int \frac{r^{5}dr}{\left( r^{4}-R^{4}\right) ^{5/4}\sqrt{1-r^{2\left(
\alpha -\beta _{3}\right) }\frac{\left( r^{4}-R^{4}\right) ^{2\beta _{3}}}{%
\left( E+Aw\right) ^{2}}vol^{2}\left( S^{3-n}\right) }}=t+t_{0},  \tag{3.15}
\end{equation}%
where $2\beta _{3}=\left( 2n-1\right) /8$.

In the both above cases the number $n$ of the flat dimensions is equal to:%
\begin{equation*}
n=0,1,2,3.
\end{equation*}%
The first and the last cases correspond to the only one Hubble parameter.

For $p=5$:%
\begin{equation}
\int \frac{r^{8}dr}{\left( r^{6}-R^{6}\right) ^{4/3}\sqrt{1-r^{2\left(
\alpha -\beta _{5}\right) }\left( r^{6}-R^{6}\right) ^{2\beta _{5}}\sigma
_{n}^{2}}}=t+t_{0},  \tag{3.16}
\end{equation}%
where $2\beta _{5}=3\left( n-2\right) /8$. The number $n$ of the flat
dimensions is equal to: $n=0,1$.

The above integrals are complicated. One can evaluated them in the limit
when the parameter $E$ goes to infinity (it means that $\sigma
_{n}\rightarrow 0$). In this case all the above integrals have simply
asymptotes: $r\sim t$. It means that the D3-brane and background $p$-branes
does not form bounded system. Thus one can notice from (3.6) that:%
\begin{equation}
\frac{H_{1}}{H_{2}}=\eta \underset{r\rightarrow \infty }{\rightarrow }%
\left\{ 
\begin{array}{cc}
\infty & p=0 \\ 
R^{2}/4 & p=1 \\ 
0 & p>1%
\end{array}%
\right.  \tag{3.17}
\end{equation}%
and $\eta $ is singular for all $p$ in $r=R$. As was mentioned above the
considered background solutions are right for $r>R$ and eq. (3.17) is valid
for big $r$. Thus the equation (3.11) for big $r$ and the background
produced by D1-branes (D-strings) takes the form:%
\begin{equation}
H_{2}^{2}\left( 1+\frac{1}{2}R^{2}\right) +\frac{1}{2}e^{-2\beta }\widetilde{%
R}=8\pi GT_{00}.  \tag{3.18}
\end{equation}%
This equation becomes the ordinary Friedmman equation with the constant
space curvature $6\widetilde{R}$ if $R=2$ which means that $H_{1}=H_{2}$.
The condition $R=2$ puts a constraint on a topological charge $g_{5}$ and a
mass $m_{5}$ of a dual D5-branes to the background D1-branes, because $R$ is
related to the charge and the mass by (2.5) and (2.6). Thus these relations
for $r_{+}=r_{-}=R$ are:%
\begin{equation}
g_{5}=\frac{3vol\left( S^{3}\right) }{\sqrt{2}\kappa }R^{3},  \tag{3.18}
\end{equation}

\begin{equation}
m_{5}=\frac{3vol\left( S^{3}\right) }{2\kappa ^{2}}R^{3}.  \tag{3.19}
\end{equation}%
Thus we get following values of $g_{5}$ and $m_{5}$:%
\begin{equation}
g_{5}=24\sqrt{2}\pi ^{2}/\kappa ,  \tag{3.15}
\end{equation}%
\begin{equation}
m_{5}=24\pi ^{2}/\kappa ^{2}.  \tag{3.16}
\end{equation}%
In the background produced by D1-branes the condition of the isotropic
expansion leads to the (3.15) and (3.16).

For $p=0$ one can see from (3.17) that the expansion of the flat dimensions
is much faster than non-flat dimensions. The second posibility for $p=0$ is
following: $H_{2}=0$ which means that the non-flat space is static. For $p>1$
the non-flat dimensions expand faster than flat or $H_{1}=0$ which means
that the flat space is static.

\section{Conclusions}

In this paper we have obtained Hublle parameters for Dk-brane embedded in
the backgrounds produced by the black p-branes. These parameters are related
to the topology of the Dk-brane: the Dk-brane is represented as the
Cartesian product of the $n$-dimensional flat space and some $(n-k)$%
-dimensional space space with the constant curvature (in our case this space
is sphere). In general case these parameters have different values. It means
that evolution from the point of view an observer fixed to the Dk-brane in
the flat and flat directions is different. The ratio of these parameters has
been obtained in explicit form for big values of $r$. This ratio is equal to
one only in one case for $p=1$ (D-strings) and for special value of $R=2$.
It means that in asymptotic region ($r\rightarrow \infty $) and for $R=2$
expansion is the same in all directions in the world-volume of the D3-brane.
In this case the mass and the topological charge are given by eqs.
(3.15-3.16 ). The above results are valid if D3-brane and background branes
does not form bounded system. It is true for sufficient big parameter $E$.
In general case the ratio $\eta $ (eq. (3.4)) depends on the position of the
D3-brane.

The considered model is an example of a toy cosmological model. The observed
isotropic expansion of our universe is realized in this model as the
condition on equality of Hubble parameters. This condition puts constraint
on the allowed masses and charges of the background D5-branes which are dual
to D1-branes. The values of the mass and\ the charge were obtained under
assumption that D3-brane has the topology of the direct product of the $n$%
-dimensional flat space and ($3-n$)-dimensional sphere. It will be
interesting relate the obtained results to the ideas presented in [20].

Form the other side in the mirage cosmology [2,3] evolution of the
world-volume is driven not only by the energy density in the world-volume
but also by the interactions with the bulk which consist of the fields and
the branes coming from string theory. The observer fixed to the world-volume
does not see the additional dimensions in which the fields and the other
branes are propagated, and interpretes these interactions as ones with the
real fields in the world-volume. The geometry (gravity) of the world-volume
is given by the induced metric and its dynamics is described by the
Friedman-like equations with the energy density modified by the contribution
coming from the bulk. In this paper the energy density ($\widetilde{T}_{00}$%
) obtained from the matter lagrangian density (eq. 2.26) is modified by the
factor $\exp \left( -2\lambda \right) $\ which depends on the position of
the D3-brane in the bulk. Thus in the equations (3.7-3.10) the energy
density $T_{00}$\ depends not only on the bare $\widetilde{T}_{00}$\ but
also on position of the D3-brane. From the world-volume perspective this
phenomena is equivalent to the introducing some fictitious matter fields
with the energy density given by $T_{00}$. In this way the obtained results
are related to the mirage cosmology.

\section{References}

[1] M. J. Duff, R. R. Khuri, J. X. Lu, \textit{String solitons}, Phys.Rep.
259, 213,(1995), hep-th/9412184 ; M. J. Duff, \textit{Supermembranes},
hep-th/9611203,

[2] A. Kehagis, E. Kiritsis, \textit{Mirage Cosmology},\ hep-th/9910174

[3] E. Kiritsis, \textit{D-branes in Standard Model building, Gravity and
Cosmology}, hep-th/0310001

[4] G. R. Dvali, G. Gabadadze, M. Porrati, Phys. Lett. \textbf{B485} 208
(2000) (hep-th/0005016); G. R. Dvali, G. Gabadadze, Phys. Rev. \textbf{D63 }%
065007 (2001) (hep-th/0008054)

[5] K. L. Panigrahi, \textit{D-brane dynamics in Dp-brane background},
Phys.Lett. B\textbf{601}, 64, (2004), hep-th/0407134

[6]\emph{\ }L. Randall, R. Sundrum, \textit{An alternative to
compactification}, Phys. Rev. Lett. 83 (1999) 4690; hep-th/9906064

[7]\emph{\ }R. Maartens, \textit{Brane-world gravity}, Living Rev.Rel. 7
(2004) 7; gr-qc/0312059

[8]\emph{\ }C. Csaki, \textit{TASI lectures on extra dimensions and branes}%
,\ hep-ph/0404096].

[9] P. Brax, C. van de Bruck and A. C. Davis, \textit{Brane world cosmology}%
, Rept.Prog.Phys. 67 (2004) 2183-2232;\ hep-th/0404011.

[10]\emph{\ }A. R. Frey, \textit{Notes on SU(3) structures in type IIB
supergravity}, JHEP 0406, 027 (2004); hep-th/0404107

[11] J. P. Gauntlett, D. Martelli, D. Waldram, \textit{Superstrings with
intrinsic torsion}, Phys. Rev. D 69, 086002 (2004); hep-th/0302158.

[12] K. Behrndt, M. Cvetic, P. Gao, \textit{General type IIB fluxes with
SU(3) structures}, Nucl. Phys. B 721, 287 (2005); hep-th/0502154.

[13] M. Gra\~{n}a, R. Minasian, M. Petrini, A. Tomasiello, \textit{%
Supersymmetric backgrounds from generalized Calabi-Yau manifolds}, JHEP
0408, 046 (2004); hep-th/0406137

[14]\emph{\ }M. Gra\~{n}a, R. Minasian, M. Petrini and A. Tomasiello, 
\textit{Generalized structures of N = 1 vacua}, hep-th/0505212.

[15] P. Chen, K. Dasgupta, K. Narayan, M. Shmakova, M. Zagermann, \textit{%
Brane Inflation, Solitons and Cosmological Solutions: I}, hep-th/0501185

[16] J. Polchinski, Phys. Rev. Lett. \textbf{75 }(1995) 4724 ; J.
Polchinski, S. Chaudhuri, C. V. Johnson, \textit{Notes on D-Branes},
hep-th/9602052

[17] G. Horowitz, A. Strominger, \textit{Black strings and p-branes,} Nucl.
Phys. \textbf{B360}, (1991) 197; M. J. Duff, J. X. Lu, \textit{Black and
super p-branes in diverse dimensions}, Nucl. Phys. \textbf{B416 }(1994) 301

[18] M. J. Duff, H. L\"{u}, C. N. Pope, \textit{The Black Branes of M-theory}%
, hep-th/9604052

[19] G. W. Gibons, K. Maeda, \textit{Black Holes and Membranes in Higher
Dimensional Theories with the Dilaton Fields,} Nucl. Phys. \textbf{B207}
(1988) 741; D. Garfinkel, G. T. Horowitz, A. Strominger, \textit{Charged
black holes in string theory},\ Phys. Rev. \textbf{D43 (}1991) 3140-3143.

[20] J.-P. Luminet, J. Weeks, A. Riazuelo, R. Lehoucq, J.-P. Uzan, \textit{%
Dodecahedral space topology as an explanation for weak wide-angle
temperature correlations in the cosmic microwave background,} Nature 425
(2003) 593, astro-ph/0310253

\end{document}